\documentclass[pdflatex,sn-nature]{sn-jnl}

\usepackage{graphicx}%
\usepackage{multirow}%
\usepackage{amsmath,amssymb,amsfonts}%
\usepackage{amsthm}%
\usepackage{mathrsfs}%
\usepackage[title]{appendix}%
\usepackage{xcolor}%
\usepackage{textcomp}%
\usepackage{manyfoot}%
\usepackage{booktabs}%
\usepackage{algorithm}%
\usepackage{algorithmicx}%
\usepackage{algpseudocode}%
\usepackage{listings}%
\usepackage{amsmath}
\usepackage{graphicx}
\usepackage{physics}
\usepackage{natbib}

\theoremstyle{thmstyleone}%

\theoremstyle{thmstyletwo}%

\theoremstyle{thmstylethree}%

\DeclareMathOperator{\sign}{sgn}

\begin{document}
\title[Emergent interactions lead to collective frustration in robotic matter]{Emergent interactions lead to collective frustration in robotic matter}

\author[1,2]{\fnm{Onurcan} \sur{Bektas}}
\equalcont{These authors contributed equally to this work.}

\author[2,3]{\fnm{Adolfo} \sur{Alsina}}
\equalcont{These authors contributed equally to this work.}

\author[1,2]{\fnm{Steffen} \sur{Rulands}*}\email{rulands@lmu.de}

\affil[1]{\orgdiv{Arnold-Sommerfeld-Center for Theoretical Physics and Center for NanoSciences}, \orgname{Ludwig-Maximilians-Universit\"at M\"unchen}, \orgaddress{\street{Theresienstr. 37}, \city{M\"unchen}, \postcode{80333}, \country{Germany}}}

\affil[2]{\orgname{Max-Planck-Institute for the Physics of Complex Systems}, \orgaddress{\street{Noethnitzer Str. 38}, \city{Dresden}, \postcode{01187}, \country{Germany}}}

\affil[3]{\orgdiv{GISC}, \orgname{Universidad Rey Juan Carlos}, \orgaddress{c/ \city{Tulipán} s/n, \postcode{28933}, \state{Móstoles}, \country{Spain}}}

\abstract{

Artificial intelligence and robotic systems are increasingly deployed as interacting collectives of learning agents. This raises the question of whether robotic matter, where many learning agents interact, shows the emergence of collective behaviour. Here we study a paradigmatic model of robotic matter and show the emergence of a range of complex, collective behaviours. Specifically, we study systems composed of stochastic interacting particles, each endowed with a deep neural network that optimises transitions based on its environment. In a one-dimensional system, we show that robotic matter exhibits complex phenomena arising from emergent interactions, including self-organisation into distinct temporal learning regimes, particle species, and long-lived frustrated states with suboptimal reward. We further identify an abrupt, density-dependent change in collective behaviour. Active matter theory suggests that this phenomenon reflects a phase transition with signatures of criticality. Our results establish robotic matter as a platform for novel non-equilibrium physics.
}

\keywords{robotic matter, emergent interactions, active matter, reinforcement learning}

\maketitle

\section{Introduction}

Recent advances in artificial intelligence (AI) have led to systems with near-human-level capabilities in some specific tasks~\cite{doi:10.1126/science.ade9097, singhal2025toward, mankowitz2023faster, webb2023emergent, he_delving_2015, dosovitskiy_learning_2017, mnih_human_level_2015}.
In commercial AI systems, multiple AI models are increasingly being deployed as interacting units, which have been shown to outperform those comprising a single neural network and to be more robust and resource-efficient~\cite{pathak_learning_2019, tang_sensory_2021, mataric_local_2000, mohammedkora23feba, shahzadlavesson13, prusadittman15aug, baker_emergent_2020}. In robotics, such systems consist of programmable robots that interact with one another and with their environment. We term such systems \emph{robotic matter}~\cite{bo2026phasesoddroboticactive}.

\begin{figure}[t]
    \centering
    \includegraphics[width=\columnwidth]{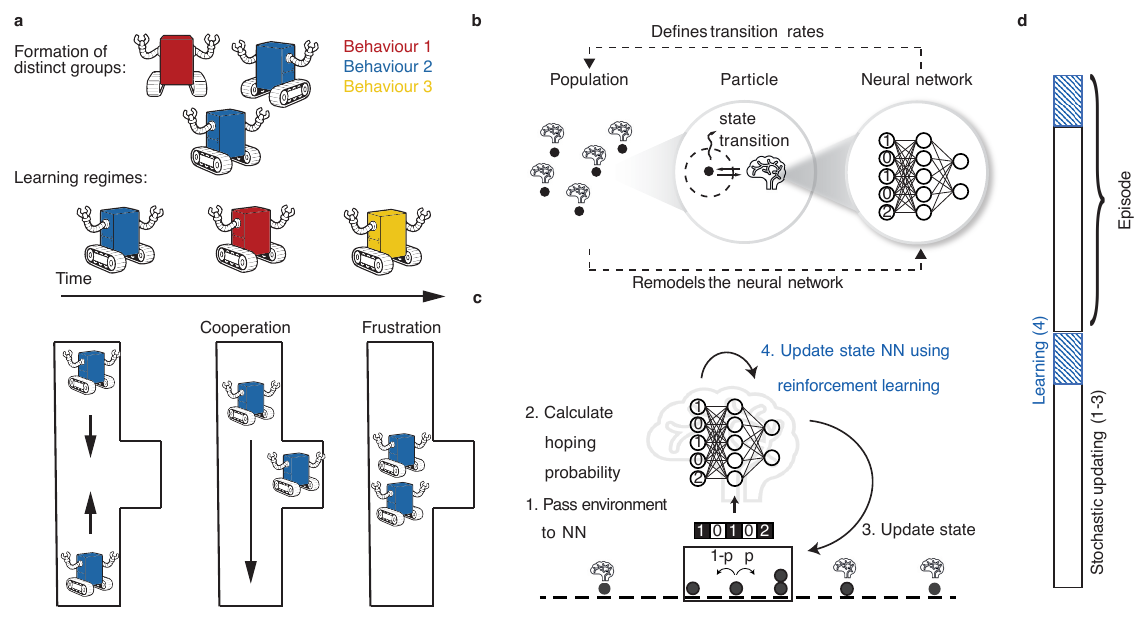}
    \caption{\textbf{Empirical phenomenology and conceptual framework of our robotic matter model.} \textbf{(a)} Four examples of counter-intuitive emergent behaviour in systems of interacting intelligent units: formation of groups of agents that learn distinct behaviours (top); temporal learning regimes, in which agents learn distinct behaviours over time (middle); cooperation, in which agents coordinate to fulfil a task (here, moving to the opposite side of the box); and frustration, in which the learned behaviour of agents leads to suboptimal outcomes. \textbf{(b)} The model paradigm comprises feedback between collective particle dynamics and the evolution of DNNs. \textbf{(c)} A minimal one-dimensional lattice-gas implementation of robotic matter. It comprises a stochastic one-dimensional lattice gas of particles whose hopping transitions are driven by deep learning. \textbf{(d)} Numerical scheme used in our simulations: we alternate between stochastic updates and neural-network training.}
    \label{fig:model}
\end{figure}

Robotic matter systems have been experimentally implemented across spatial scales and system sizes~\cite{rubensteinnagpal14auga, muinos_landincichos21mara, lowenliebchen25jan, baulinhanczyc25, dias_environmental_2023, doi:10.1126/scirobotics.adh4130, doi:10.1126/scirobotics.aax4316, bo2026phasesoddroboticactive, doi:10.1126/scirobotics.adn6035}.
Such robot collectives have been shown to exhibit a range of surprising behaviours, including suboptimal learning outcomes in reinforcement learning settings~\cite{dasguptamusolesi25mara, baker_emergent_2020}, cooperation~\cite{dasguptamusolesi25mara, baker_emergent_2020, foersterwhiteson16may, lazaridoubaroni17mara, bettini_heterogeneous_2023}, temporal learning regimes~\cite{dasguptamusolesi25mara, baker_emergent_2020, bettini_heterogeneous_2023}, or the formation of subgroups that perform different tasks~\cite{foersterwhiteson16may, lazaridoubaroni17mara, baker_emergent_2020, slavkov_morphogenesis_2018} (Fig.~\ref{fig:model}a). These behaviours are counter-intuitive because they are not directly encoded in the learning goals of the agents.

Systems comprising interacting units have long been described in statistical physics. A broad class of interacting systems out of equilibrium is active matter, in which detailed balance is microscopically broken, typically through the conversion of energy from the environment into motion~\cite{RevModPhys.85.1143, vrugt2025exactlyactivematter, cates2015motility, cates2022active, RevModPhys.88.045006, erwinfreyLesHouches2022, tan2022odd}. Robotic matter also operates far from thermal equilibrium, yet is conceptually different from active matter systems: because each unit is an independent AI system, it internally has many more degrees of freedom than the number of units in the system. Robotic matter systems therefore effectively lack microscopic symmetries. While in typical statistical physics systems one is interested in the breaking of microscopic symmetries, in robotic matter the question arises whether symmetries may emerge over time and whether these symmetries are broken again.%

Here, we take a statistical physics approach to robotic matter and show that a simple robotic matter system already exhibits highly complex spatio-temporal phenomena. In particular, we study stochastic systems in which each particle is endowed with a distinct deep neural network (DNN) that determines its stochastic state transitions as a function of its local environment.
Particles learn by analysing their trajectory over time.
Focussing on a minimal one-dimensional model, we observe a wide range of complex behaviours. Over time, increasingly complex interactions between particles evolve in a succession of temporal learning regimes.
We further observe the evolution of multiple particle species and frustration, in which the system is locked for long periods in a suboptimal state.
By using active matter theory, we show that the system exhibits a density-dependent phase transition, with signatures of criticality, and that this transition is a consequence of self-organisation mediated by emergent interactions between particles.
Our goal with this work is not to optimise learning in a particular system, but to demonstrate the broad range of emergent behaviours resulting from a single learning paradigm in such systems.

\begin{figure*}[t]
    \centering
\includegraphics[width=1\linewidth]{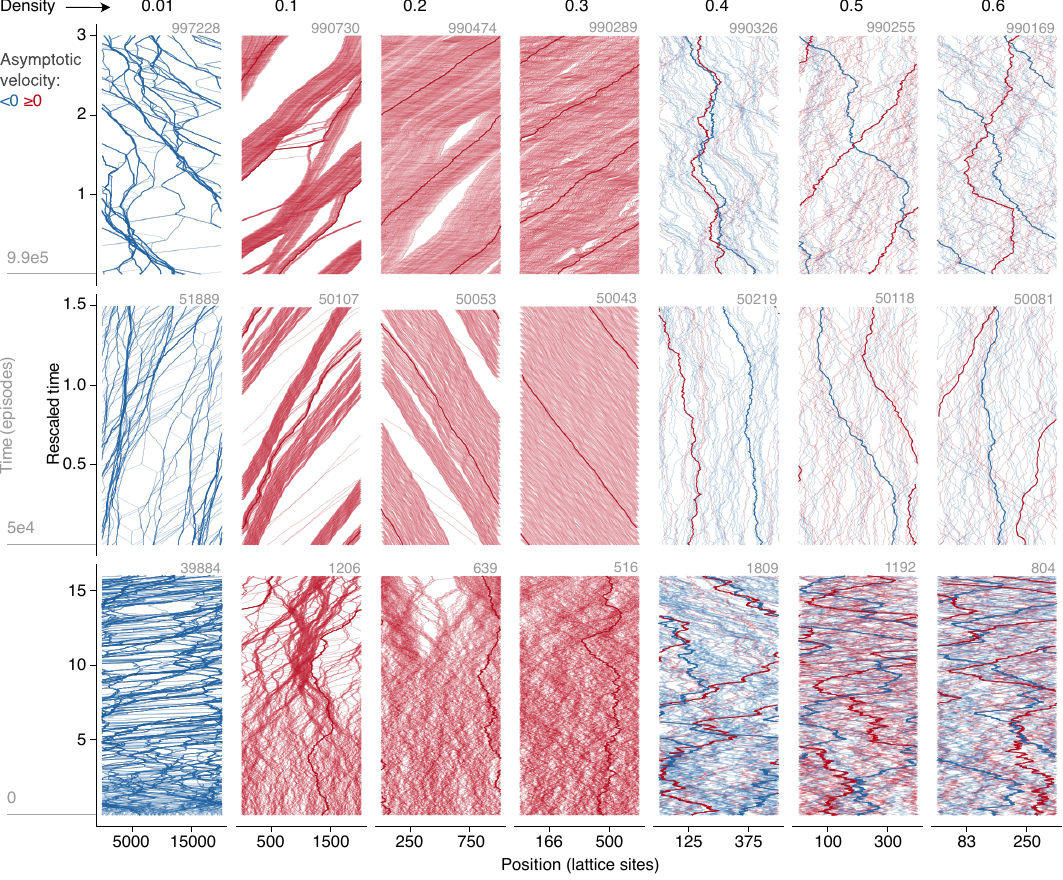}
    \caption{\textbf{Representative trajectories.} Representative trajectories showing particle positions over time in distinct temporal regimes (rows). Each column corresponds to a simulation at the specified particle density. The y-axis denotes time in episodes divided by the typical time required for a particle to cross the system once. The starting time for each row is indicated in grey on the left, and the final time is shown in the top-right corner of each panel. Line colours show the sign of the average relative velocity of each particle over the final $2\times10^5$ episodes of the simulations. Trajectories of representative particles are highlighted in bold. For densities 0.4, 0.5, and 0.6, we randomly sampled 100 particles to make individual lines distinguishable.}
    \label{fig:trajectories}
\end{figure*}

\section*{Results}
\subsection*{A framework for robotic matter}
We first define a general stochastic framework for robotic matter and then describe the specific implementation studied here. To define a paradigmatic framework, we consider a system composed of a large number of particles in which the state of each particle $i$ is described by a random variable $x_i$. %
In a description in terms of a Markov process, the dynamics is defined by transition rates between two states $\{x_i\}_{i=1,\ldots, N}$ and $\{x_i'\}_{i=1,\ldots, N}$. These transitions are generated from elementary single-particle transitions $x_i \to x_i'$. Each particle is endowed with a distinct DNN which takes the system configuration, $\{x_i\}_{i=1,\ldots, N}$, as an input and computes the probabilities of single-particle transitions as an output conditional on particle $i$ being selected for an update. The corresponding transition rates are then obtained by multiplying these conditional jump probabilities by the rate at which particle $i$ is selected. Since the output affecting a particle's behaviour depends on the DNN's input, which again contains the states of other particles, learning might lead to a situation where particles affect each other's transition rates. Such a situation would then resemble effective interactions between particles.

Sufficiently deep or wide DNNs can, in theory, approximate any continuous function up to any desired accuracy~\cite{cybenkocybenko89deca}, such that the transition rates can encode arbitrarily complex interactions. These interactions can change over time as particles learn from observing their past trajectories and update the parameters of their DNNs based on a reward function. Taken together, the particles in our model are endowed with sensors, processors and actuators~\cite{lowenliebchen25jan}. The model paradigm describes a multi-scale system, in which DNNs determine the dynamics of individual particles, and collective states on the macroscale influence the learning of the DNNs. This gives two-way feedback between macroscopic states emerging from effective interactions between particles and the dynamics of DNN parameters on the microscale (Fig.~\ref{fig:model}b).

To investigate the range of possible phenomena in such systems, we now define a specific, minimal implementation of this framework. In this implementation, the state $x_i$ of particle $i$ is defined by its position on a one-dimensional lattice of $L$ sites with periodic boundary conditions (Fig.~\ref{fig:model}c). For a given particle $i$, its environment is defined by a vector $\vec{e}_i$ whose elements give the numbers of particles present at sites in a neighbourhood of size $l$ centred on that particle. At time $t$, the particle can hop to the neighbouring site on the right with probability $p_i(\vec{e}_i,t)$ and to the neighbouring site on the left with probability $1-p_i(\vec{e}_i,t)$. The transition probability $p_i(\vec{e}_i,t)$ is determined by a feed-forward DNN that takes $\vec{e}_i$ as input. The DNN parameters associated with a given particle are dynamically updated using deep reinforcement learning on the particle's past trajectory to maximise a discounted reward. The reward function gives a large negative reward when a particle moves to an already occupied lattice site and a small positive reward otherwise (Supplementary Materials and Methods~\ref{methods:imp_smarticles}).

We used the kinetic Monte Carlo method to perform stochastic simulations of this model~\cite{gillespiegillespie76deca}. To reduce the computational cost of training the DNNs, we divided the simulation time into intervals, termed episodes (Fig.~\ref{fig:model}d). During each episode, we updated the particle positions while keeping the DNN parameters fixed. At the end of each episode, we updated the DNN parameters using reinforcement learning. Taking one episode as the unit of time, we selected each particle for an update at rate $50$, such that the corresponding continuous-time transition rates for right and left jumps were $50p_i(\vec{e}_i,t)$ and $50[1-p_i(\vec{e}_i,t)]$, respectively. Thus, each particle hopped, on average, 50 times between consecutive training events.

Two relevant approaches to training a collection of agents are multi-agent reinforcement learning (MARL) and group-agent reinforcement learning (GARL). MARL algorithms are used when multiple agents share a common environment and learn to cooperate or compete with one another~\cite{albrecht_multi_agent_nodate, lowe_multi_agent_2020, baker_emergent_2020, lazaridoubaroni17mara, duenez_guzman_statistical_2021}. These algorithms concern strategic interactions among coexisting agents in a single environment, where each agent's return depends not only on the environment but also on the evolving behaviour of the other agents.
GARL algorithms are used when each agent operates in a separate environment and communicates with other agents only to share knowledge~\cite{wu_group_agent_2023}. These algorithms concern collaborative knowledge sharing among otherwise separate agents across distinct environments. Here, we use MARL with an actor-critic temporal-difference reinforcement-learning scheme to update the DNN parameters by maximising the discounted reward calculated from particle trajectories in each episode~\cite{suttonbarto18} (see Methods). In this learning scheme, each particle has two neural networks, termed the actor and the critic; the actor generates the transition probability $p$, and the critic estimates the expected discounted return from the current state.

All figures shown here were obtained from simulations in which the number of particles was $N=200$ and the local environment size was $l=21$; the system size $L$ depended on the particle density, which is defined as $N/L$. The local environment size $l$ effectively sets the maximum range of local particle interactions. The discount factor $\gamma$, which controls the decay of contributions from times $t'>t$ to the expected reward at time $t$, was set to $0.9$. DNNs had a triangular architecture with 21 neurons in the input layer and $21$, $10$, $5$, and $2$ neurons in successive layers. We simulated up to $5\times10^6$ episodes with a learning rate of $5\times10^{-4}$, which sets the timescale of the dynamics of the DNN parameters. We also tested DNNs with $21$ neurons in each layer but observed no qualitative differences in the simulations.
To simulate dilute systems without reducing the number of particles, we increased the number of lattice sites to achieve a given particle density while keeping the environment size $l$ fixed. This ensured that the DNN sizes and training dynamics were consistent across particle densities. We take the lattice spacing as the unit of length.

In this model, detailed balance is broken locally, which can physically be interpreted as the conversion of internal energy into motion. The framework we study here therefore broadly falls into the category of active matter systems~\cite{liebchencates25jul,vrugt2025exactlyactivematter}. Robotic matter, as defined here, raises several questions that do not arise in conventional active matter systems. First, does the appearance of particles in one another's environments lead to effective interactions? Second, if such interactions exist, how do they affect learning, and can there be a physics-like description of ensuing emergent behaviour? Finally, can such effective interactions explain phenomena observed in systems of interacting robots or artificial intelligence?

\subsection*{Emergence of density-dependent frustration}
Figure~\ref{fig:trajectories} shows representative particle trajectories from simulations at different particle densities and during different time windows. Although the model defines a minimal learning rule, particles form complex spatio-temporal structures after a short transient time. These structures change qualitatively over time, indicating that the system transitions through different temporal regimes. The spatio-temporal structures also depend on particle density: at low densities, particles tend to align their velocities, whereas above a particle density of 0.3, particles move in opposite directions. We next analyse these density-dependent changes quantitatively.

\begin{figure*}[h!]
    \centering
    \includegraphics[width=1\linewidth]{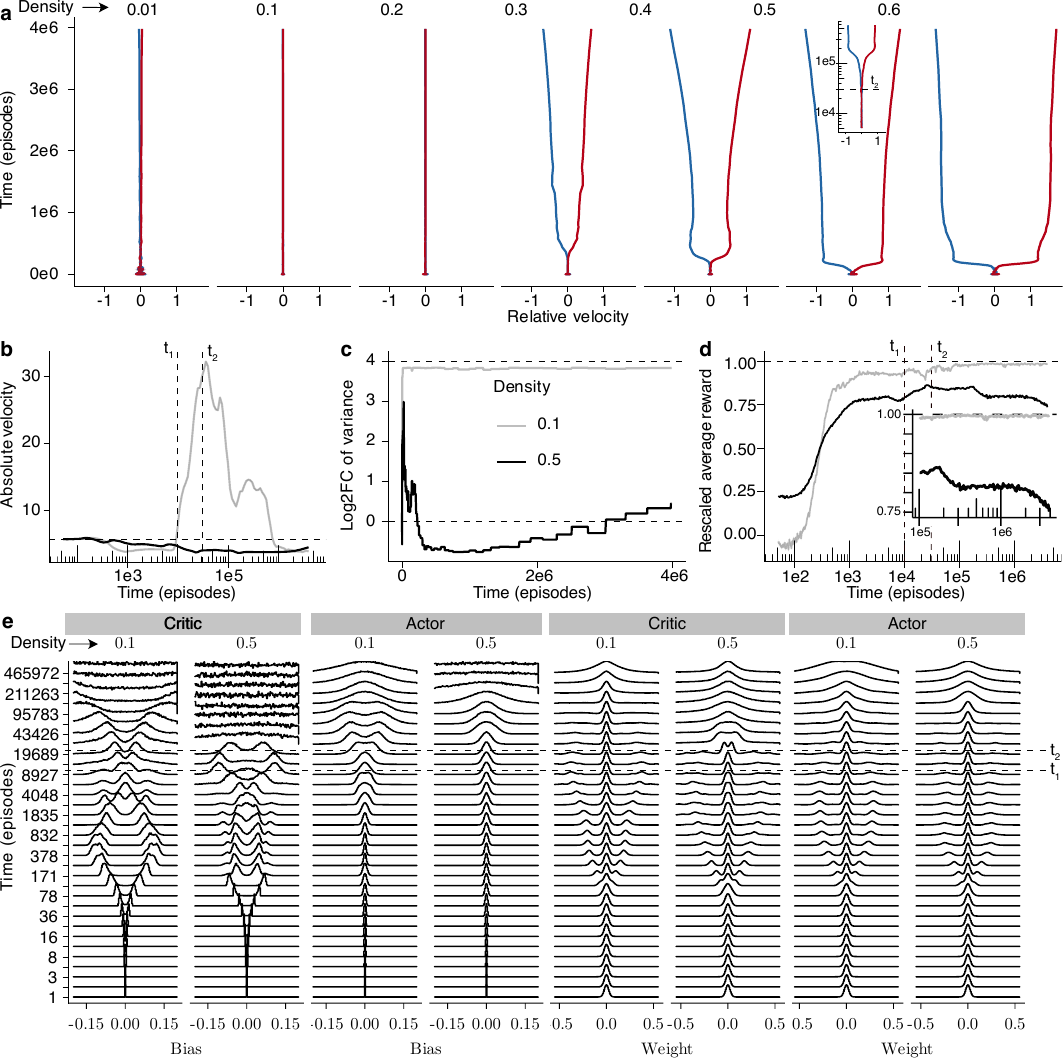}
    \caption{\textbf{Characterisation of spatio-temporal structures and time evolution of neural-network parameters.} \textbf{(a)} Average particle velocity relative to the average velocity of all particles in a simulation. Particles are grouped by their asymptotic velocity, calculated as an average over the final $2\times10^5$ episodes of the simulations. The inset shows particle density $0.5$ between episodes $5\times10^3$ and $6\times10^5$. \textbf{(b)} Average absolute velocity of all particles for two representative densities (colours defined in (c)). In this and all subsequent figures, the dashed lines on the time axis denote two characteristic times: $t_1\approx 10^4$ episodes marks the initial formation of global alignment at low densities, and $t_2\approx 3\times10^4$ episodes marks the initial formation of two groups of particles moving in opposite directions at high densities. \textbf{(c)} Base-2 logarithm of the fold change in the standard deviation of the number of right-moving particles across simulations. The lower horizontal dashed line is the standard deviation expected if the direction of motion were assigned by chance. The upper dashed line is the maximal standard deviation expected for globally aligned motion. \textbf{(d)} Average reward for two representative densities as a function of time. The inset magnifies the decrease in reward at particle density 0.5. Colours are as in (c). \textbf{(e)} Probability density functions of neural-network parameters (black lines) at logarithmically spaced time points. Density functions are shown separately for two different particle densities, for the actor and critic networks associated with each particle, and for the weights and biases in these networks.
    }
    \label{fig:velocities}
\end{figure*}

To this end, we first consider the velocity of each particle relative to the other particles. This relative velocity is defined as the difference between the instantaneous velocity of particle $i$, $v_i$, and the average velocity in the system at a given time, $\expval{v}$. We calculated the relative velocity of each particle and averaged over particles that have a relative velocity with the same sign. Figure~\ref{fig:velocities}a shows that, at densities below $0.3$, the particles have vanishing relative velocities. Thus, within each simulation run, particle motion is fully aligned, with particles moving at approximately the same velocity. To determine whether this collective velocity is non-zero and how it changes over time, we also computed the average absolute velocity (Fig.~\ref{fig:velocities}b). Global alignment emerges abruptly at a characteristic time $t_1\approx10^4$ episodes. This alignment is also consistent with the standard deviation of the number of particles moving in a given direction being close to its maximum value, as predicted by a sharply peaked bimodal distribution of asymptotic velocities across realisations (Fig.~\ref{fig:velocities}c). The average absolute velocity reaches its maximum at a time $t_2\approx3\times10^4$ episodes and decreases thereafter (Fig.~\ref{fig:velocities}b), while the relative velocity remains small. This pattern indicates collective slowing rather than a loss of alignment, consistent with the persistently high standard deviation across realisations (Fig.~\ref{fig:velocities}c). The average reward increases consistently over time at particle density 0.1, indicating that particles increasingly learn to avoid collisions (Fig.~\ref{fig:velocities}d).

By contrast, at densities of $0.3$ or greater, particles form two groups moving in opposite directions (Fig.~\ref{fig:velocities}a). However, the standard deviation in the number of particles moving in a given direction across simulations is smaller than expected under a statistically independent assignment of direction ($p-value=0.003$, F-test; Fig.~\ref{fig:velocities}c). This result implies that, in the high-density regime, particle directions are not chosen randomly but are tuned to nearly equal proportions of left- and right-moving particles at the system scale.
This fine-tuning of particle directions to a balanced 50--50 split at high densities seems counter-intuitive, given that particles receive negative rewards when they collide. Indeed, whereas the average reward at low densities increases monotonically and reaches a nearly optimal value, at high densities it first increases and then decreases at approximately the time when the two particle groups form (Fig.~\ref{fig:velocities}d).

This observation raises two questions: first, what causes the abrupt change in particle behaviour as particle density increases; and second, why is the high-density regime fine-tuned to a suboptimal reward? One possibility is that particles learn different behaviours at high and low densities, which should be reflected in the dynamics of the DNN parameters. Alternatively, the abrupt behavioural change could emerge through self-organisation arising from interactions between particles. We investigate these possibilities by examining the spatio-temporal structures, the evolution of the neural networks, and the emergent interactions between particles.

\subsection*{Spatio-temporal structures}
To quantify the evolution of spatial correlations between particle positions, we calculated the average distance to the nearest neighbour over time (Fig.~S1a). In both density regimes, the nearest-neighbour distance initially decreases, showing that particles tend to cluster. However, the average nearest-neighbour distance changes highly non-monotonically over time, reflecting distinct temporal regimes. Despite forming clusters, particles nevertheless reduce their probability of overlapping with other particles (Fig.~S1b). To further characterise the evolution of spatial structure, we calculated the pair-correlation function, which gives the average particle density as a function of distance from a given particle (Fig.~S1c). The results show that particles cluster, although sites adjacent to a given particle tend to be unoccupied. The pair-correlation function also shows that particle-particle correlations are longer-ranged in the low-density regime than in the high-density regime.

\subsection*{Time evolution of neural network parameters}
To test whether the formation of spatio-temporal structures and the density-dependent change in particle behaviour are correlated with patterns in the evolution of the DNNs, we calculated time-dependent probability density functions of the DNN parameters, namely the weights and biases. Starting from Gaussian initialisations, the DNN parameters develop different degrees of multimodality over time, indicating that they concentrate around distinct values (Fig.~\ref{fig:velocities}e).
The modality of the DNN-parameter distributions changes at the characteristic times that agree with that of changes in spatio-temporal behaviour (Fig.~\ref{fig:velocities}). Furthermore, these changes in neural-network parameters occur at times comparable to those at which the relative spatial arrangement of particles changes qualitatively, as reflected by the nearest-neighbour distance (Fig.~S1a), overlap probability (Fig.~S1b), and pair-correlation function (Fig.~S1c). This correspondence suggests the existence of distinct learning regimes in which the particles' DNNs learn different input-output relations.

\subsection*{Emergence of particle species}
To characterise interactions between particles, we considered a given environment $\vec{e}^0$, termed the base environment, and quantified the change in the transition probabilities of moving right, $p_i$, or left, $1-p_i$, when $\vec{e}^0$ was perturbed by $\Delta\vec e$. Specifically, we perturbed the base environment by adding a single particle. We then quantified interactions from the difference between the neural-network outputs for the base and perturbed environments (see Supplementary Materials):
\begin{equation}
    \nabla_{\Delta \vec{e}} p (\vec{e}^0) \approx \frac{p_i(\vec{e}^0 + \Delta \vec{e}) - p_i(\vec{e}^0)}{\left|\Delta\vec{e}\right|}. \label{eq:directed-gradient}
\end{equation}
We define $\Delta x$ as the signed displacement of the added particle from the centre of the environment. To characterise the interaction type, we multiply Eq.~\eqref{eq:directed-gradient} by $-\sign(\Delta x)$. We term the resulting quantity the \textit{repulsiveness}, $R=-\sign(\Delta x)\left[p_i(\vec{e}^0+\Delta\vec{e})-p_i(\vec{e}^0)\right]/\left|\Delta\vec{e}\right|$. Negative values of $R$ indicate attractive interactions.

\begin{figure*}[h!]
    \centering
    \includegraphics[width=1\linewidth]{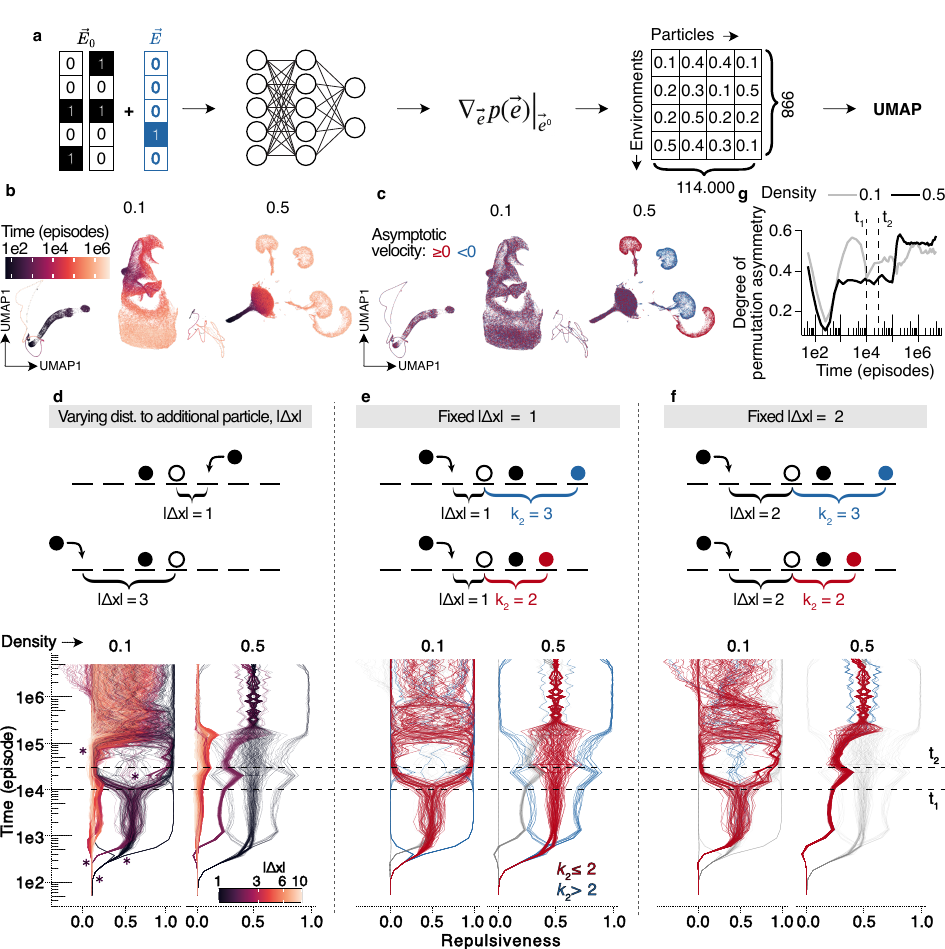}
    \caption{\textbf{Characterisation of interactions between particles.} \textbf{(a)} Schematic showing the computation of interactions. \textbf{(b)}
    Dimensionally reduced representations of interactions (UMAPs) for two representative densities (left and right). Each point represents the interaction of one particle in one environment at a given simulation time and in a given simulation run. Particles that are close in the UMAP representation have similar interaction profiles. Colours indicate \textbf{(b)} simulation time and \textbf{(c)} asymptotic velocity calculated over the final $2\times10^5$ episodes of the simulations. \textbf{(d--f)} Repulsiveness is averaged over particles and simulation runs and shown for two representative densities, 0.1 and 0.5. The stars indicate characteristic branching points of the interactions. In each panel, the upper schematics show example environment configurations, whereas the lower plots show repulsiveness for environments in which the colours indicate \textbf{(d)} the distance to the additional particle, $\Delta x$; \textbf{(e)} the distance to the second-nearest neighbour, $k_2$, when $\Delta x=1$; and \textbf{(f)} the distance to the second-nearest neighbour, $k_2$, when $\Delta x=2$. \textbf{(g)} Fraction of interactions contained in the asymmetric component over time at representative high and low densities (see Methods).}
    \label{fig:umap}
\end{figure*}

To quantify the interactions defined above for each particle, we sampled 1000 random environments from the set of environments containing at most 6 particles. At multiple time points, we passed these environments and the corresponding environments containing one additional particle to the DNN of each particle, recorded the DNN output, and calculated $R$ (Fig.~\ref{fig:umap}a). Applying this procedure to all 1000 randomly chosen environments, particles, and time points over 50 replicates per density produced roughly $10^7$ dimensions. As a first step towards quantifying particle interactions, we therefore used uniform manifold approximation and projection (UMAP), a nonlinear dimensionality-reduction technique~\cite{mcinnes2018umapsoftware}, to obtain a low-dimensional representation of the time evolution of interactions. This projection preserves local neighbourhoods; therefore, particles that are close in the projected space have similar interaction profiles. The UMAPs show the systematic evolution of particle responses to changes in their environments (Fig.~\ref{fig:umap}b). This evolution differs qualitatively between the high- and low-density regimes, indicating that particles develop different interactions in the two regimes (Fig.~\ref{fig:umap}c). The structures in the UMAPs are independent of the simulation run, showing that particles consistently develop similar sets of possible interactions across runs (Fig.~S2b).

At high densities, we observe the emergence of four clusters of particles in the projected space after a time of $2\cdot 10^5$ episodes.
This means that particles evolve into four distinct "particle species" that are distinguished for long periods by how they interact with their environments. Here, the term \textit{particle species} is used in a sense similar to its use in statistical physics or chemistry and refers to classes of particles with distinct sets of learned interactions. Two pairs of these particle species comprise particles that move ballistically in one direction, as shown in Figs.~\ref{fig:trajectories} and~\ref{fig:umap}c. Particles moving in the same direction are further separated into two species that interact differently depending on higher-order properties of their environments. We were unable to identify simple rules that distinguish between interactions of particles in the clusters corresponding to particles moving in the same direction.

\subsection*{Hierarchical evolution of interactions in learning regimes}
To gain deeper insight into these interactions, we described each environment using distances measured relative to the central particle, including the distance to the additional particle, $|\Delta x|$ (Fig.~\ref{fig:umap}d). Thus, $\Delta x$ characterises the position of the perturbation, whereas $k_1$, defined as the distance to the nearest neighbour, and $k_2$, defined as the distance to the second-nearest neighbour, characterise the pre-existing local configuration to which the DNN responds. Figure~\ref{fig:umap}d shows that interactions are predominantly repulsive and tend to become stronger over time, as expected from the choice of reward function.
Additionally, interactions tend to become longe-ranged over time, as particles develop non-zero repulsion towards other particles at increasing distances. Whereas particles at high densities interact only with particles at the nearest- and next-nearest-neighbour sites, particles at low densities interact over longer distances at late times. Figure~\ref{fig:umap}d also shows characteristic branching points after which particle responses to a given set of environments become non-homogeneous. At these time points, particles learn to distinguish an increasing number of environments.

At low densities, particles first distinguish between base environments in which $k_1=1$ and $k_2$ is either equal to or greater than 2.
At low densities, particles learn to respond to a particle at a neighbouring site ($|\Delta x|=1$) after approximately $100$ episodes (Fig.~\ref{fig:umap}e). After approximately $250$ episodes, they learn to distinguish these environments according to the position of the second-nearest neighbour, $k_2$. Specifically, particles learn to distinguish environments in which the second-nearest neighbour is two sites away from those in which it is farther away. At approximately the same time, particles learn to respond to a particle two sites away ($|\Delta x|=2$) (Fig.~\ref{fig:umap}f). After approximately $10^4$ episodes, they learn to distinguish these environments according to $k_2$. At high densities, following the emergence of nearest-neighbour repulsion, we observe a similar sequence of branching points, but at later simulation times.

A large number of environments follow similar trends, indicating that the particles' DNNs do not distinguish between these environments in terms of repulsiveness and initially do not interpret the full complexity of the environment. At characteristic times that partially overlap with the boundaries of macroscopic temporal regimes, we observe branching points after which the DNNs distinguish environments in progressively greater detail. %
Therefore, over time, DNNs of particles learn to distinguish between an increasing number of environments at characteristic time points. In between these time points, the form of interactions tends to be stable over increasingly long time periods.

Taken together, particles initially become more similar in how they interact because they all learn repulsive interactions. Later, particle groups can be distinguished by their specific responses to particular environments. This pattern may reflect the emergence and subsequent breaking of symmetries. To quantify it, we represented interactions by the repulsiveness values for different environments. Following Ref.~\cite{LU2006214}, we decomposed these interactions into a component that is symmetric with respect to permutations of particle identities and an asymmetric component. The fraction contained in the asymmetric component is zero when all particles interact in the same way and approaches one when particle-dependent deviations dominate the total interaction strength (Methods). This fraction therefore quantifies the degree of permutation asymmetry among particle identities.
Figure~\ref{fig:umap}g shows that after random initialisation, permutation symmetry emerges. Then, at time points that correspond to the branching points in Figs.~\ref{fig:umap}d-f, permutation symmetry is broken again.

\subsection*{Prediction of a phase transition}
We next address the origin of the abrupt change in the behaviour of the particles as a function of the particle density and the associated decrease in rewards.
To test the hypothesis that the abrupt change in the behaviour of particles is a consequence of self-organisation, we now develop a minimal, phenomenological model that comprises only a minimal set of interactions without explicit density dependence.

To this end, we use the framework of active-matter theory and Langevin equations. We consider the position $x_i(t)$ and velocity $v_i(t)$ of particle $i$ at a given instant in time, $t$. The evolution of its position and velocity is described by Langevin equations of the form
\begin{align}
    \dot{x}_i &= v_i, \\
    \dot{v}_i &= -\gamma(v_i)v_i + K_{i} + \sqrt{2D}\eta_i ,
\end{align}
where $\gamma(v)$ is a generalised non-equilibrium friction term that describes the intrinsic motion of the particles. Here, $K_i$ is the force exerted on particle $i$ by all other particles in its environment, $D$ is the noise amplitude, and $\eta_i$ is a Gaussian white noise.

We first identify the form of the interactions $K_i$ from our simulations. We take a coarse-grained perspective and quantify the force acting on particle $i$, $K_i(\rho,u)$, for a given average velocity field $u(x)$ and particle density field $\rho(x)$, both averaged over a small region around $x_i$. In our simulations, interactions evolve over time and are highly complex, so they cannot be readily translated into simple rules. For our argument, we identify two general, density-independent interaction components. The first follows from the predominantly repulsive particle interactions shown in Fig.~\ref{fig:umap}f, which give rise to a contribution of the form $-\pdv*{\rho(x)}{x}\eval_{x=x_i}$.

The second component arises because interactions aim to minimise not only the collision rate but also temporal correlations between collisions. Particles maximise a discounted reward of the form $G=\sum_j \gamma^j r_j$, which depends both on the instantaneous collision rate and on temporal correlations between collisions. Specifically, the expected discounted reward $\langle G\rangle$ increases as these correlations decrease (Supplementary Materials and Methods).
The minimisation of the autocorrelation time is empirically supported by our simulations (Fig.~S2a). The autocorrelation time is proportional to the time that a pair of particles spends in one another's vicinity, such that particles minimise a function proportional to $1/|v-u|$. This leads to a force given by the derivative of this term with respect to $v$, $\sign(v-u)/|v-u|^2$. Because the correlation scales with particle density, the second contribution to the interaction term takes the form $\rho\sign(v-u)/|v-u|^2$.

\begin{figure*}[h!]
    \centering
    \includegraphics[width=\textwidth]{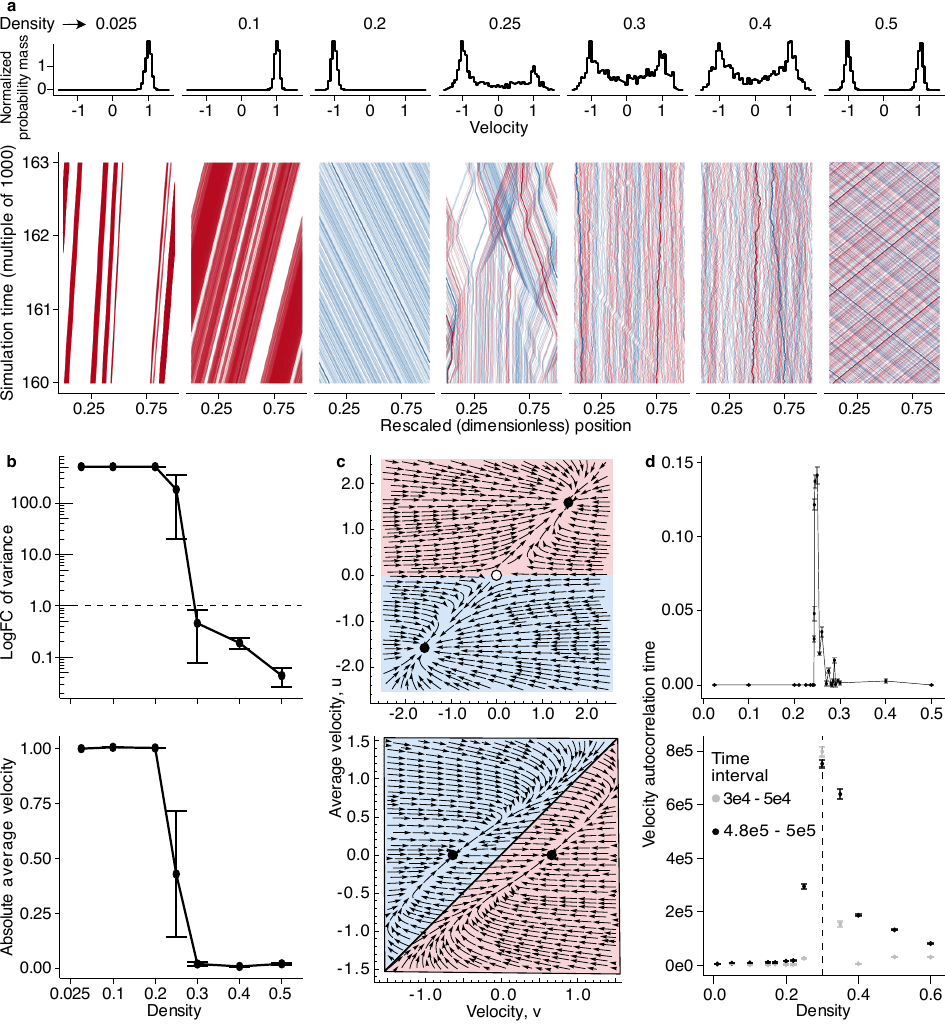}
    \caption{\textbf{Prediction of a density-dependent phase transition using a Langevin description.} \textbf{(a)} (top) Distribution of particle velocities at the final time point of Langevin simulations of Eqs.~\eqref{eq:langevin-x} and~\eqref{eq:langevin-v}. (bottom) Representative trajectories from Langevin simulations showing particle positions at the specified particle density (see Supplementary Materials and Methods for details). The colours show the sign of the average particle velocity over the final 100 time points (red, positive; blue, negative). At densities greater than 0.2, we randomly sampled 256 particles to make individual lines distinguishable. \textbf{(b)} (top) Log-fold change in the variance of the number of right-moving particles in Langevin simulations as a function of density. Error bars show the standard error of the mean. The horizontal dashed line is the variance expected if the direction of motion were assigned by chance. (bottom) Asymptotic absolute average velocity as a function of particle density. \textbf{(c)} (top) Phase portrait of Eqs. S13 and S14 at particle density 0.1. (bottom) Phase portrait of Eqs. S13 and S14 at particle density 0.5. The colours show the sign of the test-particle velocity in the Hartree approximation (see Supplementary Materials and Methods). \textbf{(d)} (top) Velocity autocorrelation time obtained from simulations of Eqs.~\eqref{eq:langevin-x} and~\eqref{eq:langevin-v}. (bottom) Velocity autocorrelation time obtained from simulations of the full model in Fig.~\ref{fig:model}b (see Supplementary Materials and Methods for details). The dashed line indicates particle density 0.3.}
    \label{fig:theory}
\end{figure*}

We now define the active-friction term $\gamma(v)$. Because our system is out of equilibrium, the fluctuation-dissipation theorem does not hold, and friction is therefore not coupled to the noise amplitude. Instead, this term comprises two parts: one describing directed motion and the other describing friction. A phenomenological expression that captures both effects is the Rayleigh--Helmholtz friction~\cite{erdmannschweitzer00may}, which has been used to describe a variety of active-matter systems. It has the form $\gamma(v)=-\alpha+\beta v^2$. The inertial term describes the persistent particle motion arising from non-zero bias parameters that develop early in our simulations (Fig.~\ref{fig:velocities}e).
Taken together, in non-dimensional form, the time evolution of the velocity and position of a given particle is
\begin{align}
    \dot x &= v, \label{eq:langevin-x} \\
    \dot v &= v - v^3 + s \frac{\rho \sign{(v - u)}}{(v - u)^2}  - r \pdv{\rho}{x} + \sqrt{B} \eta, \label{eq:langevin-v}
\end{align}
where we have dropped the index $i$. We have non-dimensionalised the Langevin equations by rescaling time and space by $t' = \alpha t$ and $x' = x\sqrt{\alpha \beta}$, respectively, and removed the apostrophes to simplify the notation (see Supplementary Materials and Methods~\ref{ap:nondim_langevin}). The dimensionless parameters $s = \zeta \beta^2/\alpha^2$ and $r = \delta \sqrt{\beta^3/\alpha}$ quantify the strength of interactions due to the minimisation of correlations and repulsion relative to the effect of self-propulsion, respectively. $B = 2 D \beta/\alpha^2$ is the dimensionless noise strength. $\zeta,\delta$ are some constants with appropriate units.

In this model, particle density fluctuations are weak at high densities, so we expect the term that minimises correlations to dominate. Intuitively, this term maximises the difference between the particle velocity $v$ and the average velocity $u$, which could produce the global split observed in our simulations. At low densities, the correlation-minimising term is negligible, and we expect particle repulsion to dominate. %

To test this prediction numerically, we performed stochastic simulations of Eqs.~\eqref{eq:langevin-x} and~\eqref{eq:langevin-v} using the Euler--Maruyama scheme (see Supplementary Materials and Methods). We found global particle alignment at low densities and a split into groups moving in opposite directions at high densities (Fig.~\ref{fig:theory}a). At high density, the variance in the number of particles moving in the positive direction across simulations was again smaller than expected under the null model (t-test, $p=6.16\times10^{-7}$), suggesting collective tuning of the sizes of the two groups (top section of Fig.~\ref{fig:theory}b). Furthermore, at a density of approximately $0.25$, the asymptotic absolute average velocity decreases abruptly from $1$ to $0$ (bottom section of Fig.~\ref{fig:theory}b). These observations suggest that the abrupt change in particle behaviour shown in Fig.~\ref{fig:velocities}a is a phase transition arising from collective behaviour mediated by effective particle interactions.

To investigate this behaviour more quantitatively, we analytically derived the co-evolution of the velocity of a particle, $v$, and the average velocity of all particles, $\bar{u}$~\cite{romanczuk_collective_2010} (see Supplementary Materials and Methods). %
The resulting dynamics are summarised in phase portraits (Fig.~\ref{fig:theory}c). The phase portraits show attractors corresponding to asymptotic dynamical regimes of Eqs.~\eqref{eq:langevin-x} and~\eqref{eq:langevin-v}. At low densities, we find two attractors with $v=\bar{u}$, corresponding to states in which the particle velocity aligns with the average system velocity. At high densities, these attractors lie on the horizontal line $\bar{u}=0$. Thus, the phenomenological model predicts that individual velocities approach specific positive or negative values under the constraint of a vanishing global velocity. Because the line $\bar{u}=0$ is stable, feedback mechanisms suppress fluctuations around it. This corresponds to the observed state with precisely tuned equal proportions of particles moving in either direction.

Taken together, our analysis suggests that the abrupt density-dependent change in particle behaviour is an emergent phenomenon resulting from effective interactions between particles. It is therefore a phase transition of the type observed in statistical-physics systems. At all densities, particle interactions have macroscopic effects. In the high-density regime, particles optimise their reward locally without predicting the suboptimal macroscopic consequences of their behaviour. This resembles a learning analogue of frustration phenomena in disordered systems or the tragedy of the commons in the social sciences.
This analysis raises the question of whether the phase transition resembles known transitions in non-equilibrium statistical physics. To investigate this question, we calculated autocorrelation functions of individual-particle velocities at different densities in both the Langevin simulations and simulations of the full model comprising intelligent agents. In both cases, the autocorrelation time diverges at the particle density at which the phase transition occurs (Fig.~\ref{fig:theory}d). Diverging correlations are characteristic of critical points and therefore suggest critical behaviour. Further characterisation of the phase transition would require the computation of critical exponents, which is currently not feasible for the intelligent-agent model. Notably, the transition occurs at a particle density for which the range of each particle's repulsive interactions multiplied by the number of particles equals the lattice size; that is, the effective volume fraction equals 1. This observation could indicate a relation to jamming transitions.

\section*{Discussion}
Taken together, we studied a paradigmatic model of groups of learning and intelligent agents, termed robotic matter, which are increasingly used in artificial intelligence and robotics. Our extensive simulations reveal a rich phenomenology, including emergent phenomena that lead to a phase transition, the evolution of multiple particle species, temporal learning regimes, and a phenomenon akin to frustration in which particles adopt suboptimal strategies. Our analytical and numerical work shows that the key phenomenology results from self-organisation and should therefore not depend on details of the numerical implementation, provided that the relation between fundamental timescales and learning convergence is maintained. Indeed, we tested whether our results depended on the size and structure of the DNNs, the learning rate, or the discount factor and found no qualitative changes.

These phenomena mirror empirical observations made in much more complex robot collectives or ensembles of neural networks. Because these observations have been made in highly complex or commercial systems, their interpretation has remained qualitative and has not been linked to an emergent phenomenon. Our work provides an explanation for these observations in terms of fundamental self-organisation processes.

First, these observations include temporal learning regimes in which neural networks learn different behaviours~\cite{baker_emergent_2020, dasguptamusolesi25mara, bettini_heterogeneous_2023}. Our work clarifies the emergence of such regimes in robotic and artificial-intelligence systems. In our model, these learning regimes arise as non-equilibrium metastable states defined by the combination of learned interactions and their macroscopic consequences at the system level. They are destabilised as agents progress through a learning hierarchy in which increasingly complex environments are interpreted.
Second, both virtual and mechanical robot collectives have been shown to form distinct groups of agents, such as leaders and followers, or groups that develop distinct behaviours~\cite{foersterwhiteson16may, lazaridoubaroni17mara, baker_emergent_2020, balchbalch00jun, slavkov_morphogenesis_2018, liabu_mostafa04dec}. We show that particle species can form through self-organisation without the prior existence of different learning tasks. Finally, groups of learning agents can encounter ``deadlock'' situations~\cite{dasguptamusolesi25mara, baker_emergent_2020}, resembling the physical concept of frustration in disordered systems. Our work clarifies how self-organisation and emergence lead to suboptimal behaviour because agents do not learn to predict the macroscopic consequences of their local optimisation. Taken together, our work implies that macroscopic collective behaviour can influence learning drastically and unpredictably.

The active-matter system studied here differs qualitatively from those typically considered in active-matter theory: because each particle has many intrinsic degrees of freedom and the number of these degrees of freedom exceeds the number of particles that can typically be simulated, the system effectively lacks microscopic symmetries. Whereas statistical physics typically focuses on the breaking of microscopic symmetries, here the question is whether symmetries emerge through learning and are subsequently broken. In our model, symmetries first emerge through the development of generic particle interactions. These emergent symmetries are then broken, as evidenced by the formation of particle species. Robotic matter, as defined here, therefore provides a testing ground for new concepts in non-equilibrium physics.

\section*{Methods}
\subsection*{Implementation of the stochastic simulations}\label{methods:imp_smarticles}
We used the Gillespie algorithm to simulate particle jumps~\cite{gillespiegillespie76deca}. During each Gillespie event, we first drew the waiting time $\Delta t$ from an exponential distribution with rate $50N$, where $N$ is the total number of particles. We set the duration of each episode to $1$ (Fig.~\ref{fig:model}d) and then chose one particle uniformly at random. This choice fixes the unit of time such that one unit of simulation time corresponds, on average, to $50$ jumps per particle per unit of time. The episode duration was chosen to be long enough to accumulate sufficient training data but shorter than the typical timescales of global system evolution. The selected particle read the local occupancy state vector, $\vec{e}$, for the lattice region of length $l$ centred on its current location. This vector was then passed to the particle's actor, which output the probabilities $p_j(\vec{e},t)$ and $1-p_j(\vec{e},t)$ that particle $j$ would jump to the right and left, respectively, at time $t$. The jump was executed regardless of the occupancy of the target site. After each jump, the particle received a scalar reinforcement signal, termed the reward, that was used only to update the parameters of the selected particle's actor and critic. The reward was $+1$ for a jump to an empty site and $-100$ otherwise. The reward scale is proportional to the learning rate. The asymmetry between positive and negative rewards ensures that the learning signal dominates despite the stochasticity of the presented environments.
During each episode, we recorded (i) the environments provided as input to the actors, (ii) the direction of each jump, and (iii) the reward obtained after each jump.
At the end of each episode, the neural networks of all particles were trained using the recorded triplets $(\text{observed environment}, \text{jump direction}, \text{reward received})$. For each particle $i$, only its own trajectory was used to train its actor and critic. In accordance with temporal-difference reinforcement learning, we trained the actor using the loss
$L_j^{\textbf{actor}} = -\sum_{i=1}^{n_j-1} \delta_j (t_{i,j}) \log{\pi_j} (t_{i,j})$,
where $n_j$ is the number of jumps made by particle $j$ during the episode and $\delta_j(t_{i,j}):=R_j(t_{i,j})+\gamma V_j(t_{i+1,j})-V_j(t_{i,j})$. Here, $\pi_j(t_{i,j})$, $R_j(t_{i,j})$, and $V_j(t_{i,j})$ denote the jump probability, received reward, and critic-predicted value, respectively, at jump time
$t_{i,j}$. The set $\{t_{i,j}\mid i=1,\ldots,n_j\}$ contains the times at which particle $j$ jumped. For rightward jumps, $\pi_j(t_{i,j})=p_j(e_j(t_{i,j}),t_{i,j})$; otherwise, $\pi_j(t_{i,j})=1-p_j(e_j(t_{i,j}),t_{i,j})$. Here, $e_j(t_{i,j})$ is the state of the local environment observed by particle $j$ at time $t_{i,j}$. We trained the critic using the loss $L_j^{\textbf{critic}}=\sum_{i=1}^{n_j-1}\delta_j(t_{i,j})^2/(n_j-1)$.
During training, gradients of these loss functions were used to update the actor and critic parameters with the Adam optimiser.

\subsection*{Numerical solution of the Langevin equations}\label{methods:imp_langevin}
We used the explicit Euler--Maruyama scheme to integrate the non-dimensionalised Langevin equation. We tested systems of $N=200$ and $N=1024$ particles using a time increment $\delta t=0.001/\tau_t$ for a total time $T=1\times10^{5}/\tau_t$, where $\tau_t=1/\alpha$ is the characteristic timescale. The system size was determined from $L=N/\langle\rho\rangle$, where $\langle\rho\rangle=\rho_0\times\tau_l$ is the average non-dimensional particle density, $\rho_0$ is the particle density, and $\tau_l=1/\sqrt{\alpha\beta}$ is the characteristic length scale. We used periodic boundary conditions and set $\alpha=16$, $\beta=1$, $\delta=50$, $\zeta=75$, and $D=2.5$. We set the interaction radius to $1/\tau_l$ when calculating the local particle density and velocity fields $\rho(x)$ and $u(x)$. For $N=200$, we ran 15--24 replicates per particle density, whereas for $N=1024$, we ran 2 replicates per particle density because of limited GPU availability. To prevent divergences when numerically integrating the singular term $\frac{\sign(v'-u')}{(v'-u')^2}$, we regularised it using the $\tanh$ function.

\subsection*{Non-dimensionalisation of the Langevin equation} \label{ap:nondim_langevin}
To non-dimensionalise the Langevin equations, we rescaled time and position as
\begin{equation}
    t' := \alpha t ,\quad x' := \sqrt{\alpha \beta} x
\end{equation}
This leads to a rescaling of velocities as
\begin{equation}
    v' := \sqrt{\frac{\beta}{\alpha}} v, \quad u' := \sqrt{\frac{\beta}{\alpha}} u ,
\end{equation}
and of the particle density as
\begin{equation*}
    \rho' := \frac{\rho}{\sqrt{\alpha \beta}}.
    \end{equation*}
The rescaled noise term reads
\begin{equation*}
            \eta' := \frac{\eta}{\sqrt{\alpha}}.
    \end{equation*}
\newpage
Using these normalised variables, we obtain dimensionless Langevin equations of the form
\begin{align}
    \frac{\dd x'}{\dd t'} &= v',\\
    \frac{\dd v'}{\dd t'} &= v' - v'^3 + \zeta \frac{\beta^2}{\alpha^2} \frac{\rho' \sign{(v' - u')}}{(v' - u')^2}\\
    &- \delta \sqrt{\frac{\beta^3}{\alpha}} \partial_{x'} \rho' + \sqrt{\frac{2 D \beta}{\alpha^2}} \eta'.
\end{align}

\subsection*{Calculation of the velocity autocorrelation function} \label{ap:methods_autocorr}
We used the batch-means method to calculate the velocity autocorrelation time of individual particles in the full model comprising intelligent agents (Fig.~\ref{fig:theory}d(top))~\cite{thompson_comparison_2010}. For individual particles in the many-body Langevin model, we calculated the connected correlation function
\begin{equation}
    C_i(t, t') = \langle v_i (t)  v_i(t') \rangle - \langle v_i(t) \rangle \langle v_i(t') \rangle
\end{equation}
for each particle $i$. We then calculated the autocorrelation time using its definition
\begin{equation}
    \tau_i = \int_{t'}^{t' + T} \mathrm{d}t\, C_i(t, t') / C_i(t, t').
\end{equation}
Here, $T$ was set to $500$.

\subsection*{Quantification of symmetries in particle interactions}
In statistical physics, microscopic symmetries are usually defined as invariances of the microscopic dynamical rules under a group of transformations. For a system of identical particles, one such symmetry is permutation symmetry: relabelling particles should not change the microscopic laws of motion. In an interacting particle system with fixed interactions, this symmetry is usually built into the model by construction, since all particles obey the same transition rules and interact through the same set of interactions.

The robotic matter system studied here is different. Each particle is endowed with its own neural network, and the transition probability of particle $i$ is determined by the parameters of that particle's network. Therefore, even if all networks have the same architecture and are initialised from the same distribution, the microscopic transition rules are not constrained to be identical for different particles. The system therefore does not possess exact microscopic permutation symmetry by construction. Over time, however, learning can either drive the particles towards a common set of interactions, so that permutation symmetry emerges, or drive them towards qualitatively distinct sets of interactions, corresponding to the formation of particle species discussed in the main text.

As a result, the degree to which particle interactions possess permutation symmetry over particle indices measures their similarity. To formalise this idea, we borrow a method from dynamical-systems theory and decompose the state of a system into longitudinal and transverse components relative to a manifold~\cite{LU2006214}. Here, the manifold is the subspace of interaction matrices with identical rows. We provide a brief description of this approach below.

We consider permutation symmetry in the interaction space. If such symmetry emerges, different particles respond identically to the same perturbation of their environment. Conversely, this symmetry is absent when particles acquire qualitatively distinct interactions, for example when they form species distinguished by their responses to probe environments. This situation is observed in the UMAP representation of the interactions, where particles form long-lived clusters corresponding to distinct interaction sets.

To quantify this idea, we first construct an interaction matrix $R(t)\in\mathbb{R}^{N\times M}$, where $N$ is the number of particles and $M$ is the number of probe environments. The element $R_{i\alpha}(t)$ denotes the response of particle $i$ to probe environment $\alpha$ at time $t$. In practice, $R_{i\alpha}$ is the repulsiveness defined from the change in right-jump probability upon adding a particle to the local environment. Thus, each row $R_i(t)$ is the interaction vector of one particle across all probe environments.

The fully permutation-symmetric subspace consists of all matrices whose rows are identical,
$$
\mathcal{S}=\left\{S \in \mathbb{R}^{N \times M}: S_{i \alpha}=\mu_\alpha \text { for all } i\right\} .
$$
A matrix in this subspace describes one type of interaction $\mu_\alpha$ shared by all particles. Therefore, the component of $R(t)$ inside $\mathcal{S}$ is the part of the learned interaction structure that is common to all particles.

We obtain this component by orthogonally projecting $R(t)$ onto $\mathcal{S}$. With respect to the Frobenius inner product, the closest matrix in $\mathcal{S}$ is obtained by replacing every row of $R(t)$ by the average row,
$$
\bar{R}_\alpha(t)=\frac{1}{N} \sum_{i=1}^N R_{i \alpha}(t) .
$$
Equivalently, $R_{i \alpha}^{\mathrm{sym}}(t)=\bar{R}_\alpha(t)$. This projection can be written as $R^{\mathrm{sym}}(t)=P R(t),$
where $P=\frac{1}{N}\mathbf{1}\mathbf{1}^{\top}$ is the orthogonal projector onto the particle-independent mode. The remaining component,
$$
R^{\mathrm{asym}}(t)=R(t)-R^{\mathrm{sym}}(t)=(I-P) R(t),
$$
contains the deviations of individual particles from the common interaction law. It is therefore the component transverse to the permutation-symmetric subspace. This gives the orthogonal decomposition
$$
R(t)=R^{\mathrm{sym}}(t)+R^{\mathrm{asym}}(t),
$$
with
$$
\|R(t)\|_F^2=\left\|R^{\mathrm{sym}}(t)\right\|_F^2+\left\|R^{\mathrm{asym}}(t)\right\|_F^2,
$$
where $\|.\|_F$ denotes the Frobenius norm. We then define the symmetric interaction fraction
$$
m_{\mathrm{sym}}(t)=\frac{\left\|R^{\mathrm{sym}}(t)\right\|_F^2}{\|R(t)\|_F^2}=\frac{\|P R(t)\|_F^2}{\|R(t)\|_F^2}.
$$
The complementary quantity
$$
\Delta_{\mathrm{asym}}(t)=1-m_{\mathrm{sym}}(t)=\frac{\|(I-P) R(t)\|_F^2}{\|R(t)\|_F^2}
$$
measures the fraction of interactions that is carried by particle-dependent deviations from the common set of interactions, termed the \textit{permutation asymmetry fraction}. Thus, $\Delta_{\text {asym}}=0$ corresponds to complete permutation symmetry of the learned interactions.

\section*{Acknowledgements}
We thank Alexander Ziepke, Stefano Bo, Frank J\"ulicher, and the entire Rulands group for helpful discussions.

\section*{Funding}
This research was supported by the Center for Nano Science (CeNS).

\section*{Author contributions}
A.A. and O.B. contributed equally. A.A., O.B., and S.R. conceptualised the work. A.A. and S.R. supervised the work. O.B. wrote the simulation code and analysed data. O.B. performed analytical calculations. S.R. acquired funding. O.B. and S.R. wrote the initial draft of the manuscript. All authors edited and approved the final version of the manuscript.

\section*{Data availability}
The data are available from the authors upon reasonable request.

\bibliography{references}

@article{he_delving_2015,
  title = {Delving Deep into Rectifiers: Surpassing Human-Level Performance on ImageNet Classification},
  author = {He, Kaiming and Zhang, Xiangyu and Ren, Shaoqing and Sun, Jian},
  journal = {Proceedings of the IEEE International Conference on Computer Vision},
  year = {2015},
  pages = {1026--1034}
}

@article{mnih_human_level_2015,
  title = {Human-Level Control through Deep Reinforcement Learning},
  author = {Mnih, Volodymyr and Kavukcuoglu, Koray and Silver, David and Rusu, Andrei A. and Veness, Joel and Bellemare, Marc G. and Graves, Alex and Riedmiller, Martin and Fidjeland, Andreas K. and Ostrovski, Georg and Petersen, Stig and Beattie, Charles and Sadik, Amir and Antonoglou, Ioannis and King, Helen and Kumaran, Dharshan and Wierstra, Daan and Legg, Shane and Hassabis, Demis},
  year = {2015},
  month = feb,
  journal = {Nature},
  volume = {518},
  number = {7540},
  pages = {529--533},
  issn = {1476-4687},
  doi = {10.1038/nature14236}
}

@article{
dosovitskiy_learning_2017,
title={Learning to Act by Predicting the Future},
author={Alexey Dosovitskiy and Vladlen Koltun},
journal={International Conference on Learning Representations},
year={2017}
}

@article{pathak_learning_2019,
author = {Pathak, Deepak and Lu, Chris and Darrell, Trevor and Isola, Phillip and Efros, Alexei A.},
title = {Learning to control self-assembling morphologies: a study of generalization via modularity},
year = {2019},
journal = {Proceedings of the 33rd International Conference on Neural Information Processing Systems},
articleno = {206},
numpages = {11}
}

@article{tang_sensory_2021,
 author = {Tang, Yujin and Ha, David},
 journal = {Advances in Neural Information Processing Systems},
 editor = {M. Ranzato and A. Beygelzimer and Y. Dauphin and P.S. Liang and J. Wortman Vaughan},
 pages = {22574--22587},
 publisher = {Curran Associates, Inc.},
 title = {The Sensory Neuron as a Transformer: Permutation-Invariant Neural Networks for Reinforcement Learning},
 volume = {34},
 year = {2021}
}

@book{mataric_local_2000,
  title = {From {{Local Interactions}} to {{Collective Intelligence}}},
  booktitle = {Prerational {{Intelligence}}: {{Adaptive Behavior}} and {{Intelligent Systems Without Symbols}} and {{Logic}}, {{Volume}} 1, {{Volume}} 2 {{Prerational Intelligence}}: {{Interdisciplinary Perspectives}} on the {{Behavior}} of {{Natural}} and {{Artificial Systems}}, {{Volume}} 3},
  author = {Matari{\'c}, Maja J.},
  editor = {Cruse, Holk and Dean, Jeffrey and Ritter, Helge},
  year = {2000},
  pages = {988--998},
  publisher = {Springer Netherlands},
  address = {Dordrecht},
  doi = {10.1007/978-94-010-0870-9_61}
}

@article{mohammedkora23feba,
    title = {A comprehensive review on ensemble deep learning: {Opportunities} and challenges},
    volume = {35},
    issn = {1319-1578},
    shorttitle = {A comprehensive review on ensemble deep learning},
    doi = {10.1016/j.jksuci.2023.01.014},
    number = {2},
    journal = {Journal of King Saud University - Computer and Information Sciences},
    author = {Mohammed, Ammar and Kora, Rania},
    month = feb,
    year = {2023},
    pages = {757--774},
}

@article{shahzadlavesson13,
    title = {Comparative {Analysis} of {Voting} {Schemes} for {Ensemble}-based {Malware} {Detection}},
    volume = {4},
    number = {1},
    journal = {Journal of Wireless Mobile Networks, Ubiquitous Computing, and Dependable Applications},
    author = {Shahzad, Raja Khurram and Lavesson, Niklas},
    year = {2013},
    pages = {98--117}
}

@article{prusadittman15aug,
    title = {Using {Ensemble} {Learners} to {Improve} {Classifier} {Performance} on {Tweet} {Sentiment} {Data}},
    doi = {10.1109/IRI.2015.49},
    journal = {2015 {IEEE} {International} {Conference} on {Information} {Reuse} and {Integration}},
    author = {Prusa, Joseph and Khoshgoftaar, Taghi M. and Dittman, Daivd J.},
    month = aug,
    year = {2015},
    pages = {252--257},
}

@article{
baker_emergent_2020,
title={Emergent Tool Use From Multi-Agent Autocurricula},
author={Bowen Baker and Ingmar Kanitscheider and Todor Markov and Yi Wu and Glenn Powell and Bob McGrew and Igor Mordatch},
journal={International Conference on Learning Representations},
year={2020}
}

@article{rubensteinnagpal14auga,
  title = {Programmable Self-Assembly in a Thousand-Robot Swarm},
  author = {Rubenstein, Michael and Cornejo, Alejandro and Nagpal, Radhika},
  year = {2014},
  month = aug,
  journal = {Science},
  volume = {345},
  number = {6198},
  pages = {795--799},
  publisher = {American Association for the Advancement of Science},
  doi = {10.1126/science.1254295}
}

@inbook{lowenliebchen25jan,
author="L{\"o}wen, Hartmut
and Liebchen, Benno",
editor="te Vrugt, Michael",
chapter="Towards Intelligent Active Particles",
title="Artificial Intelligence and Intelligent Matter: Nanoscience, Soft Matter, Philosophy",
year="2026",
publisher="Springer Nature Switzerland",
address="Cham",
pages="257--271",
isbn="978-3-032-04129-6",
doi="10.1007/978-3-032-04129-6_13"
}

@article{baulinhanczyc25,
  title = {Intelligent Soft Matter: {{Towards}} Embodied Intelligence},
  shorttitle = {Intelligent Soft Matter},
  author = {Baulin, Vladimir A. and Giacometti, Achille and Fedosov, Dmitry A. and Ebbens, Stephen and {Varela-Rosales}, Nydia R. and Feliu, Neus and Chowdhury, Mithun and Hu, Minghan and F{\"u}chslin, Rudolf and Dijkstra, Marjolein and Mussel, Matan and Van Roij, Ren{\'e} and Xie, Dong and Tzanov, Vassil and Zu, Mengjie and {Hidalgo-Caballero}, Samuel and Yuan, Ye and Cocconi, Luca and Ghim, Cheol-Min and {Cottin-Bizonne}, C{\'e}cile and Miguel, M. Carmen and Esplandiu, Maria Jose and Simmchen, Juliane and Parak, Wolfgang J. and Werner, Marco and Gompper, Gerhard and Hanczyc, Martin M.},
  year = {2025},
  journal = {Soft Matter},
  volume = {21},
  number = {21},
  pages = {4129--4145},
  issn = {1744-683X, 1744-6848},
  doi = {10.1039/D5SM00174A}
}

@article{
muinos_landincichos21mara,
author = {S. Muiños-Landin  and A. Fischer  and V. Holubec  and F. Cichos },
title = {Reinforcement learning with artificial microswimmers},
journal = {Science Robotics},
volume = {6},
number = {52},
pages = {eabd9285},
year = {2021},
doi = {10.1126/scirobotics.abd9285}
}

@article{dias_environmental_2023,
  title = {Environmental Memory Boosts Group Formation of Clueless Individuals},
  author = {Dias, Crist{\'o}v{\~a}o S. and Trivedi, Manish and Volpe, Giovanni and Ara{\'u}jo, Nuno A. M. and Volpe, Giorgio},
  year = {2023},
  month = nov,
  journal = {Nature Communications},
  volume = {14},
  number = {1},
  pages = {7324},
  issn = {2041-1723},
  doi = {10.1038/s41467-023-43099-0}
}

@article{dasguptamusolesi25mara,
  title = {Investigating the Impact of Direct Punishment on the Emergence of Cooperation in Multi-Agent Reinforcement Learning Systems},
  author = {Dasgupta, Nayana and Musolesi, Mirco},
  year = 2025,
  month = mar,
  journal = {Autonomous Agents and Multi-Agent Systems},
  volume = {39},
  number = {1},
  pages = {19},
  issn = {1573-7454}
}

@article{foersterwhiteson16may,
author = {Foerster, Jakob N. and Assael, Yannis M. and de Freitas, Nando and Whiteson, Shimon},
title = {Learning to communicate with Deep multi-agent reinforcement learning},
year = {2016},
isbn = {9781510838819},
publisher = {Curran Associates Inc.},
address = {Red Hook, NY, USA},
journal = {Proceedings of the 30th International Conference on Neural Information Processing Systems},
pages = {2145–2153},
numpages = {9},
location = {Barcelona, Spain},
series = {NIPS'16}
}

@article{
lazaridoubaroni17mara,
title={Multi-Agent Cooperation and the Emergence of (Natural) Language},
author={Angeliki Lazaridou and Alexander Peysakhovich and Marco Baroni},
journal={International Conference on Learning Representations},
year={2017}
}

@article{bettini_heterogeneous_2023,
author = {Bettini, Matteo and Shankar, Ajay and Prorok, Amanda},
title = {Heterogeneous Multi-Robot Reinforcement Learning},
year = {2023},
isbn = {9781450394321},
publisher = {International Foundation for Autonomous Agents and Multiagent Systems},
address = {Richland, SC},
journal = {Proceedings of the 2023 International Conference on Autonomous Agents and Multiagent Systems},
pages = {1485–1494},
numpages = {10},
location = {London, United Kingdom},
series = {AAMAS '23}
}

@article{balchbalch00jun,
  title = {Hierarchic {{Social Entropy}}: {{An Information Theoretic Measure}} of {{Robot Group Diversity}}},
  shorttitle = {Hierarchic {{Social Entropy}}},
  author = {Balch, Tucker},
  year = 2000,
  month = jun,
  journal = {Autonomous Robots},
  volume = {8},
  number = {3},
  pages = {209--238},
  issn = {1573-7527}
}

@article{slavkov_morphogenesis_2018,
  title = {Morphogenesis in Robot Swarms},
  author = {Slavkov, I. and {Carrillo-Zapata}, D. and Carranza, N. and Diego, X. and Jansson, F. and Kaandorp, J. and Hauert, S. and Sharpe, J.},
  year = {2018},
  month = dec,
  journal = {Science Robotics},
  volume = {3},
  number = {25},
  pages = {eaau9178},
  doi = {10.1126/scirobotics.aau9178}
}

@article{liabu_mostafa04dec,
  title = {Learning and {{Measuring Specialization}} in {{Collaborative Swarm Systems}}},
  author = {Li, Ling and Martinoli, Alcherio and {Abu-Mostafa}, Yaser S.},
  year = 2004,
  month = dec,
  journal = {Adaptive Behavior},
  volume = {12},
  number = {3-4},
  pages = {199--212},
  publisher = {SAGE Publications Ltd STM},
  issn = {1059-7123}
}

@article{cybenkocybenko89deca,
  title = {Approximation by Superpositions of a Sigmoidal Function},
  author = {Cybenko, G.},
  year = {1989},
  month = dec,
  journal = {Mathematics of Control, Signals and Systems},
  volume = {2},
  number = {4},
  pages = {303--314},
  issn = {1435-568X},
  doi = {10.1007/BF02551274}
}

@article{gillespiegillespie76deca,
  title = {A General Method for Numerically Simulating the Stochastic Time Evolution of Coupled Chemical Reactions},
  author = {Gillespie, Daniel T},
  year = {1976},
  month = dec,
  journal = {Journal of Computational Physics},
  volume = {22},
  number = {4},
  pages = {403--434},
  issn = {0021-9991},
  doi = {10.1016/0021-9991(76)90041-3}
}

@book{albrecht_multi_agent_nodate,
  title = {Multi-Agent Reinforcement Learning: {{Foundations}} and Modern Approaches},
  author = {Albrecht, Stefano V. and Christianos, Filippos and Sch{\"a}fer, Lukas},
  year = {2024},
  publisher = {MIT Press}
}

@article{lowe_multi_agent_2020,
 author = {Lowe, Ryan and WU, YI and Tamar, Aviv and Harb, Jean and Pieter Abbeel, OpenAI and Mordatch, Igor},
 journal = {Advances in Neural Information Processing Systems},
 editor = {I. Guyon and U. Von Luxburg and S. Bengio and H. Wallach and R. Fergus and S. Vishwanathan and R. Garnett},
 pages = {},
 publisher = {Curran Associates, Inc.},
 title = {Multi-Agent Actor-Critic for Mixed Cooperative-Competitive Environments},
 volume = {30},
 year = {2017}
}

@misc{duenez_guzman_statistical_2021,
      title={Statistical discrimination in learning agents}, 
      author={Edgar A. Duéñez-Guzmán and Kevin R. McKee and Yiran Mao and Ben Coppin and Silvia Chiappa and Alexander Sasha Vezhnevets and Michiel A. Bakker and Yoram Bachrach and Suzanne Sadedin and William Isaac and Karl Tuyls and Joel Z. Leibo},
      year={2021},
      eprint={2110.11404},
      archivePrefix={arXiv},
      primaryClass={cs.LG}
}

@article{wu_group_agent_2023,
author="Wu, Kaiyue
and Zeng, Xiao-Jun",
editor={"Iliadis, Lazaros
and Papaleonidas, Antonios
and Angelov, Plamen
and Jayne, Chrisina"},
title="Group-Agent Reinforcement Learning",
journal="Artificial Neural Networks and Machine Learning",
year="2023",
publisher="Springer Nature Switzerland",
address="Cham",
pages="37--48",
isbn="978-3-031-44223-0"
}

@book{suttonbarto18,
  title = {Reinforcement Learning: {{An}} Introduction},
  shorttitle = {Reinforcement Learning},
  author = {Sutton, Richard S. and Barto, Andrew G.},
  year = {2018},
  series = {Adaptive Computation and Machine Learning Series},
  edition = {Second edition},
  publisher = {The MIT Press},
  address = {Cambridge, Massachusetts},
  isbn = {978-0-262-03924-6}
}

@misc{liebchencates25jul,
      title={What exactly is 'active matter'?}, 
      author={Michael te Vrugt and Benno Liebchen and Michael E. Cates},
      year={2025},
      eprint={2507.21621},
      archivePrefix={arXiv},
      primaryClass={cond-mat.soft} 
}

@article{erdmannschweitzer00may,
  title = {Brownian Particles Far from Equilibrium},
  author = {Erdmann, U. and Ebeling, W. and {Schimansky-Geier}, L. and Schweitzer, F.},
  year = {2000},
  month = may,
  journal = {The European Physical Journal B - Condensed Matter and Complex Systems},
  volume = {15},
  number = {1},
  pages = {105--113},
  issn = {1434-6036},
  doi = {10.1007/s100510051104}
}

@article{romanczuk_collective_2010,
  title = {Collective Motion of Active {{Brownian}} Particles in One Dimension},
  author = {Romanczuk, P. and Erdmann, U.},
  year = {2010},
  month = sep,
  journal = {The European Physical Journal Special Topics},
  volume = {187},
  number = {1},
  pages = {127--134},
  issn = {1951-6401},
  doi = {10.1140/epjst/e2010-01277-0}
}

@misc{thompson_comparison_2010,
      title={A Comparison of Methods for Computing Autocorrelation Time}, 
      author={Madeleine B. Thompson},
      year={2010},
      eprint={1011.0175},
      archivePrefix={arXiv},
      primaryClass={stat.CO}
}

@misc{vrugt2025exactlyactivematter,
      title={What exactly is 'active matter'?}, 
      author={Michael te Vrugt and Benno Liebchen and Michael E. Cates},
      year={2025},
      eprint={2507.21621},
      archivePrefix={arXiv},
      primaryClass={cond-mat.soft}
}

@book{cates2022active,
  title={Active Matter and Nonequilibrium Statistical Physics},
  author={Cates, ME and Tailleur, J},
  year={2022},
  publisher={Oxford University Press Oxford}
}

@article{cates2015motility,
  title={Motility-induced phase separation},
  author={Cates, Michael E and Tailleur, Julien},
  journal={Annu. Rev. Condens. Matter Phys.},
  volume={6},
  number={1},
  pages={219--244},
  year={2015},
  publisher={Annual Reviews}
}

@article{RevModPhys.88.045006,
  title = {Active particles in complex and crowded environments},
  author = {Bechinger, Clemens and Di Leonardo, Roberto and L\"owen, Hartmut and Reichhardt, Charles and Volpe, Giorgio and Volpe, Giovanni},
  journal = {Rev. Mod. Phys.},
  volume = {88},
  issue = {4},
  pages = {045006},
  numpages = {50},
  year = {2016},
  month = {Nov},
  publisher = {American Physical Society},
  doi = {10.1103/RevModPhys.88.045006}
}

@inbook{erwinfreyLesHouches2022,
    author="Frey, Erwin and Brauns, Fridtjof",
    editor = {Tailleur, Julien and Gompper, Gerhard and Marchetti, M. Cristina and Yeomans, Julia M. and Salomon, Christophe},
    chapter = {Self-organization of protein patterns},
    title = {Active Matter and Nonequilibrium Statistical Physics: Lecture Notes of the Les Houches Summer School: Volume 112, September 2018},
    publisher = {Oxford University Press},
    year = {2022},
    month = {11},
    isbn = {9780192858313},
    doi = {10.1093/oso/9780192858313.001.0001}
}

@article{RevModPhys.85.1143,
  title = {Hydrodynamics of soft active matter},
  author = {Marchetti, M. C. and Joanny, J. F. and Ramaswamy, S. and Liverpool, T. B. and Prost, J. and Rao, Madan and Simha, R. Aditi},
  journal = {Rev. Mod. Phys.},
  volume = {85},
  issue = {3},
  pages = {1143--1189},
  numpages = {0},
  year = {2013},
  month = {Jul},
  publisher = {American Physical Society},
  doi = {10.1103/RevModPhys.85.1143}
}

@article{tan2022odd,
  title={Odd dynamics of living chiral crystals},
  author={Tan, Tzer Han and Mietke, Alexander and Li, Junang and Chen, Yuchao and Higinbotham, Hugh and Foster, Peter J and Gokhale, Shreyas and Dunkel, J{\"o}rn and Fakhri, Nikta},
  journal={Nature},
  volume={607},
  number={7918},
  pages={287--293},
  year={2022},
  publisher={Nature Publishing Group UK London}
}

@article{
doi:10.1126/scirobotics.adh4130,
author = {Baudouin Saintyves  and Matthew Spenko  and Heinrich M. Jaeger },
title = {A self-organizing robotic aggregate using solid and liquid-like collective states},
journal = {Science Robotics},
volume = {9},
number = {86},
pages = {eadh4130},
year = {2024},
doi = {10.1126/scirobotics.adh4130}
}

@article{
doi:10.1126/scirobotics.aax4316,
author = {William Savoie  and Thomas A. Berrueta  and Zachary Jackson  and Ana Pervan  and Ross Warkentin  and Shengkai Li  and Todd D. Murphey  and Kurt Wiesenfeld  and Daniel I. Goldman },
title = {A robot made of robots: Emergent transport and control of a smarticle ensemble},
journal = {Science Robotics},
volume = {4},
number = {34},
pages = {eaax4316},
year = {2019},
doi = {10.1126/scirobotics.aax4316}
}

@misc{bo2026phasesoddroboticactive,
      title={Three phases of odd robotic active matter}, 
      author={Fan Bo and Shiqi Liu and Zenghong He and Wyatt Joyce and Gregor Leech and Kiet Tran and Keilan Ramirez and Nicholas Boechler and Nicholas Gravish and Hongbo Zhao and Tzer Han Tan},
      year={2026},
      eprint={2603.09897},
      archivePrefix={arXiv},
      primaryClass={cond-mat.soft}
}

@article{
doi:10.1126/science.ade9097,
author = {Anton Bakhtin and Noam Brown  and Emily Dinan  and Gabriele Farina  and Colin Flaherty  and Daniel Fried  and Andrew Goff  and Jonathan Gray  and Hengyuan Hu  and Athul Paul Jacob  and Mojtaba Komeili  and Karthik Konath  and Minae Kwon  and Adam Lerer  and Mike Lewis  and Alexander H. Miller  and Sasha Mitts  and Adithya Renduchintala  and Stephen Roller  and Dirk Rowe  and Weiyan Shi  and Joe Spisak  and Alexander Wei  and David Wu  and Hugh Zhang  and Markus Zijlstra },
title = {Human-level play in the game of Diplomacy by combining language models with strategic reasoning},
journal = {Science},
volume = {378},
number = {6624},
pages = {1067-1074},
year = {2022},
doi = {10.1126/science.ade9097}
}

@article{singhal2025toward,
  title={Toward expert-level medical question answering with large language models},
  author={Singhal, Karan and Tu, Tao and Gottweis, Juraj and Sayres, Rory and Wulczyn, Ellery and Amin, Mohamed and Hou, Le and Clark, Kevin and Pfohl, Stephen R and Cole-Lewis, Heather and others},
  journal={Nature medicine},
  volume={31},
  number={3},
  pages={943--950},
  year={2025},
  publisher={Nature Publishing Group US New York}
}

@article{mankowitz2023faster,
  title={Faster sorting algorithms discovered using deep reinforcement learning},
  author={Mankowitz, Daniel J and Michi, Andrea and Zhernov, Anton and Gelmi, Marco and Selvi, Marco and Paduraru, Cosmin and Leurent, Edouard and Iqbal, Shariq and Lespiau, Jean-Baptiste and Ahern, Alex and others},
  journal={Nature},
  volume={618},
  number={7964},
  pages={257--263},
  year={2023},
  publisher={Nature Publishing Group UK London}
}

@article{webb2023emergent,
  title={Emergent analogical reasoning in large language models},
  author={Webb, Taylor and Holyoak, Keith J and Lu, Hongjing},
  journal={Nature Human Behaviour},
  volume={7},
  number={9},
  pages={1526--1541},
  year={2023},
  publisher={Nature Publishing Group UK London}
}

@article{
doi:10.1126/scirobotics.adn6035,
author = {Daniel I. Goldman  and D. Zeb Rocklin },
title = {Robot swarms meet soft matter physics},
journal = {Science Robotics},
volume = {9},
number = {86},
pages = {eadn6035},
year = {2024},
doi = {10.1126/scirobotics.adn6035}}

@article{LU2006214,
title = {New approach to synchronization analysis of linearly coupled ordinary differential systems},
journal = {Physica D: Nonlinear Phenomena},
volume = {213},
number = {2},
pages = {214-230},
year = {2006},
issn = {0167-2789},
doi = {https://doi.org/10.1016/j.physd.2005.11.009},
author = {Wenlian Lu and Tianping Chen}
}

@article{mcinnes2018umapsoftware,
  title={UMAP: Uniform Manifold Approximation and Projection},
  author={McInnes, Leland and Healy, John and Saul, Nathaniel and Grossberger, Lukas},
  journal={Journal of Open Source Software},
  volume={3},
  number={29},
  pages={861},
  year={2018},
  doi={10.21105/joss.00861}
}

\end{document}